\documentclass[aps,prd,twocolumn,nofootinbib,showpacs,amsmath,amssymb,amsfonts,floatfix,twoside]{revtex4-1}

\usepackage[utf8]{inputenc}
\usepackage{subfig}
\usepackage{graphicx,bm,hyperref,enumitem,array,dcolumn}
\hypersetup{breaklinks=true}

\makeatletter
\def\p@paragraph{\thesection\,\thesubsection\,}
\makeatother

\renewcommand{\Re}{\operatorname{Re}}
\renewcommand{\Im}{\operatorname{Im}}

\newcommand{\kfm}{k_{\mathrm{fm}}}
\newcommand{\kmb}{k_{\mathrm{mb}}}
\newcommand{\qbkfm}{\dfrac{qb}{\kfm}}
\newcommand{\dqq}{dq\, q}
\newcommand{\dbb}{db\, b}
\newcommand{\bess}{J_0 \left(\qbkfm\right)}

\newcommand{\sigmat}{\sigma_{\mathrm{tot}}}
\newcommand{\Ginel}{G_{\mathrm{inel}}}

\begin{document}

\title{Impact-parameter analysis of the new TOTEM pp data at 13~TeV: the black disk limit excess.}

\author{A. Alkin}
\affiliation{%
Bogolyubov Institute for Theoretical Physics, Metrologichna 14b, Kiev, UA-03680, Ukraine.
}%
\author{O. Kovalenko }
\affiliation{%
National Centre for Nuclear Research, ul. Hoża 69, PL-00-681 Warsaw, Poland.
}%
\author{E. Martynov}
\affiliation{%
Bogolyubov Institute for Theoretical Physics, Metrologichna 14b, Kiev, UA-03680, Ukraine.
}%
\author{S.M. Troshin}
\affiliation{%
 NRC ``Kurchatov Institute''--IHEP,
 Protvino, Moscow Region, 142281, Russia.
}%

\date{\today}

\begin{abstract}
We revisit a discussion on the impact-parameter dependence of proton-proton elastic scattering amplitude with improved uncertainty calculation. 
This analysis allows to reveal the asymptotic properties of hadron interactions. 
New data indicates that the impact-parameter elastic scattering amplitude is slightly above the black disk limit at 13~TeV c.m.s. energy of the LHC reaching a value of $\Im H(s,0) = 0.512\pm 0.001\ \text{(sys+stat)} \pm 0.004\ \text{(norm)}$ confirming that black disk limit is violated at current collision energy, however it was not exceeded at 7~TeV.
The growth trend of the impact-parameter amplitude imaginary part, extrapolated from previous and new data, indicates that it is unlikely that the amplitude is close to saturation. 
New analysis is consistent with smooth energy evolution of the elastic scattering amplitude and supersedes the earlier conclusion on the black disk limit excess observed at 7~TeV. 
\end{abstract}

\maketitle

\section{Introduction}
Due to a rapid growth of experimentally available proton-proton collision energy we are provided with a unique opportunity to test the general asymptotic properties of hadronic collisions. 
In particular, asymptotic at $s\to\infty$ ratio of elastic to total cross-section $\sigma_{\mathrm{el}} / \sigmat$ can range from $1/2$, the so called \emph{black disk limit}, corresponding to maximal elastic unitarity contribution from inelastic channel, to $1$, corresponding to maximal partial amplitudes allowed by unitarity.
This ratio is directly related to the value of elastic amplitude at zero impact parameter.

In 1980  U.~Amaldi and K. R.~Schubert \cite{Amaldi:1979kd} suggested an approach to reconstruct hadronic elastic amplitude dependence on impact parameter from differential cross-section data. 
It was first applied to available data at $\sqrt{s} = 546$~GeV by T.~Fearnley in 1985 \cite{Fearnley:1985uy}. 
Only in 2014 the modified method was applied to the latest elastic pp scattering data at $\sqrt{s} = 7$~TeV \cite{Alkin:2014rfa}.
The analysis yielded an intriguing result, showing, for the first time, that black disk limit is violated.
Such feature of the elastic pp amplitude also casts doubt on eikonal approximations related to this limit.
While at 7~TeV the effect was small, it was expected to increase with collision energy.
This observation inspired a number of publications, analyzing phenomenological and theoretical consequences of elastic amplitude behavior, such as the recent talk by V.A.~Petrov and A.P.~Samokhin \cite{Petrov:2018wlv} and a few other examples \cite{Dremin:2013qua,Fagundes:2015lba,RuizArriola:2016ihz}. 
However, the original analysis was flawed, it was later found that at 7~TeV $\Im H(0)$ is still below $0.5$ (see paragraph \ref{par:ImH}).
Finally, in 2018 TOTEM collaboration has presented pp differential cross-section data at 13~TeV \cite{Ravera:2018,*Nemes:2018} that allows us to revise previous observation. 
In this paper we present the results of elastic amplitude impact parameter dependence reconstruction from new TOTEM data using an updated analysis, showing that at the current energy black disk limit violation can be seen.

\section{\label{sec:analysis}Analysis}
The general approach follows that of the previous publication \cite{Alkin:2014rfa}. 
Here we provide a short summary and a description of updated uncertainty calculation method. 
The starting point is the impact-parameter representation of hadronic amplitude $H(s,b)$ defined by the transformation of the elastic scattering amplitude $A(s,t)$
\begin{align}
H(s,b)&=\dfrac{1}{8\pi s}\int_{0}^{\infty}\dqq \bess A(s,t),\label{eq:b-transform-to} \\
A(s,t)&=8\pi s\int_{0}^{\infty}\dbb \bess H(b,s),\label{eq:b-transform-from}
\end{align}
where $\kfm = 0.1973269718$~GeV$\,$fm, $J_0(x)$ is a Bessel function of the first kind and $q^{2} \equiv -t$. 
Normalization of $A(s,t)$ is chosen to be
\begin{equation}\label{eq:normalization}
\sigmat=\frac{\kmb}{s} \Im A(s,0), \quad \frac{d\sigma}{dt}=\frac{\kmb}{16\pi s^{2}}\left|A(s,t)\right|^{2}
\end{equation}
where $\kmb = 0.389379338$~mb$\,$GeV$^{2}$ and $\sigmat$ is the value of total pp cross-section in millibarns.

The elastic amplitude dependence on momentum transfer $t$ was first parameterized, for the experimentally available region $8\times 10^{-4}\text{~GeV}^2 \leq \left| t \right| \leq 3.83\text{~GeV}^2$, using the functional form introduced in previous analysis \cite{Alkin:2014rfa}.
It is referred to as a \emph{standard} parameterization
\begin{multline}\label{eq:standard-parameterization}
A(t)=s\left\{ i\alpha \left [A_{1}e^{\alpha b_{1} t/2}+(1-A_{1})e^{\alpha b_{2} t/2}\right ]\right .\\
\left . -iA_{2}e^{b_{3}t/2}-A_{2}\rho (1-t/\tau)^{-4}\right\}\!,
\end{multline}
where
\begin{equation}
\alpha=(1-i\rho)\left( \sigmat / \kmb + A_{2}\right)
\end{equation}
and experimental value is used for $\rho$ parameter, $\rho \equiv \Re A(s,0) / \Im A(s,0)$.
Additionally, the exponential parameterization was used in a general form
\begin{equation}\label{eq:generic-parameterization}
A(t) = s \times \left( \Re A + i\Im A \right), 
\end{equation}
where real and imaginary parts are
\begin{align}
\label{eq:real-part-alternatives}
\Re A & = A_4 e^{b_4 \xi} + A_5 e^{b_5 \xi},\ \xi \equiv 3m_{\pi} - \sqrt{9m_{\pi}^2 - t} \\
\label{eq:imaginary-part-alternatives}
\Im A & = A_1 e^{b_1 t} + A_2 e^{b_2 t} + A_3 e^{b_3 t} + A_{p}\left(1 - t/\tau\right)^{-4}.
\end{align}
Two variants are considered, referred to as \emph{exponential (3+1)} and \emph{exponential (3+2)} parameterizations, that correspond to cases where either $A_5$ or $A_p$ is fixed at zero.
Relations between unknown constants $A_i$ were determined by requiring parameterizations to automatically satisfy equations $\Im A(s,0)=s\sigmat/k_{\mathrm{mb}}$ and $\Re A(s,0)/\Im A(s,0)=\rho$ for arbitrary parameter values
\begin{equation}\label{eq:constraints}
\begin{array}{ll}
A_3 &= \sigmat / \kmb - A_1 - A_2 - A_p, \\
A_4 & = \rho \sigmat / \kmb - A_5.
\end{array}
\end{equation}
Parameters were fitted to $d\sigma / d t$ data using relation \eqref{eq:normalization} with $\sigmat$ and $\rho$ considered free parameters, limited according to their experimental values and uncertainties.
 
\paragraph{Data set and low-$|t|$ region description.} Two TOTEM data sets were combined to have the widest possible $t$ range covered. 
The \mbox{low-$|t|$} data set \cite{Antchev:2017yns} is fully compatible with the most recent results for larger momentum transfer \cite{Ravera:2018,*Nemes:2018}.
To improve the quality of the parameterization at low values of $|t|$, a Coulomb term in the simplest form \cite{West:1968du} was added to all three parameterization variants 
\begin{multline}
A_{C}(t) = \frac{s}{t} \frac{8\pi \alpha_{\mathrm{EM}}}{\left(1 + t/t_0\right)^4} \\
\times \exp \left\{ - i \left[\alpha_{\mathrm{EM}} \left(\gamma + \ln \frac{B \left|t\right|}{2} \right) \right]\right\},
\end{multline}
where $\alpha_{\mathrm{EM}} \approx 0.007297$ is the fine-structure constant, $t_0 = 0.71$~GeV$^2$, $\gamma = 0.577$ and $B = 20.4$~GeV$^{-2}$ is the first cone slope of differential cross-section determined by exponential fit (the value coincides with the one obtained by TOTEM \cite{Antchev:2017yns}). 
We emphasize that this form of Coulomb term was chosen for its simplicity, rather than physical essence, in order to achieve better data description at low $|t|$ and, consequently, improve the overall fit quality.
An actual analysis of the Coulomb-nuclear interference region is performed by TOTEM Collaboration \cite{Antchev:2017yns}.
The normalization-related systematic uncertainty was excluded from both data sets for fitting and was propagated to the final result separately.
\begin{figure}[tb]
	\includegraphics[width=\linewidth]{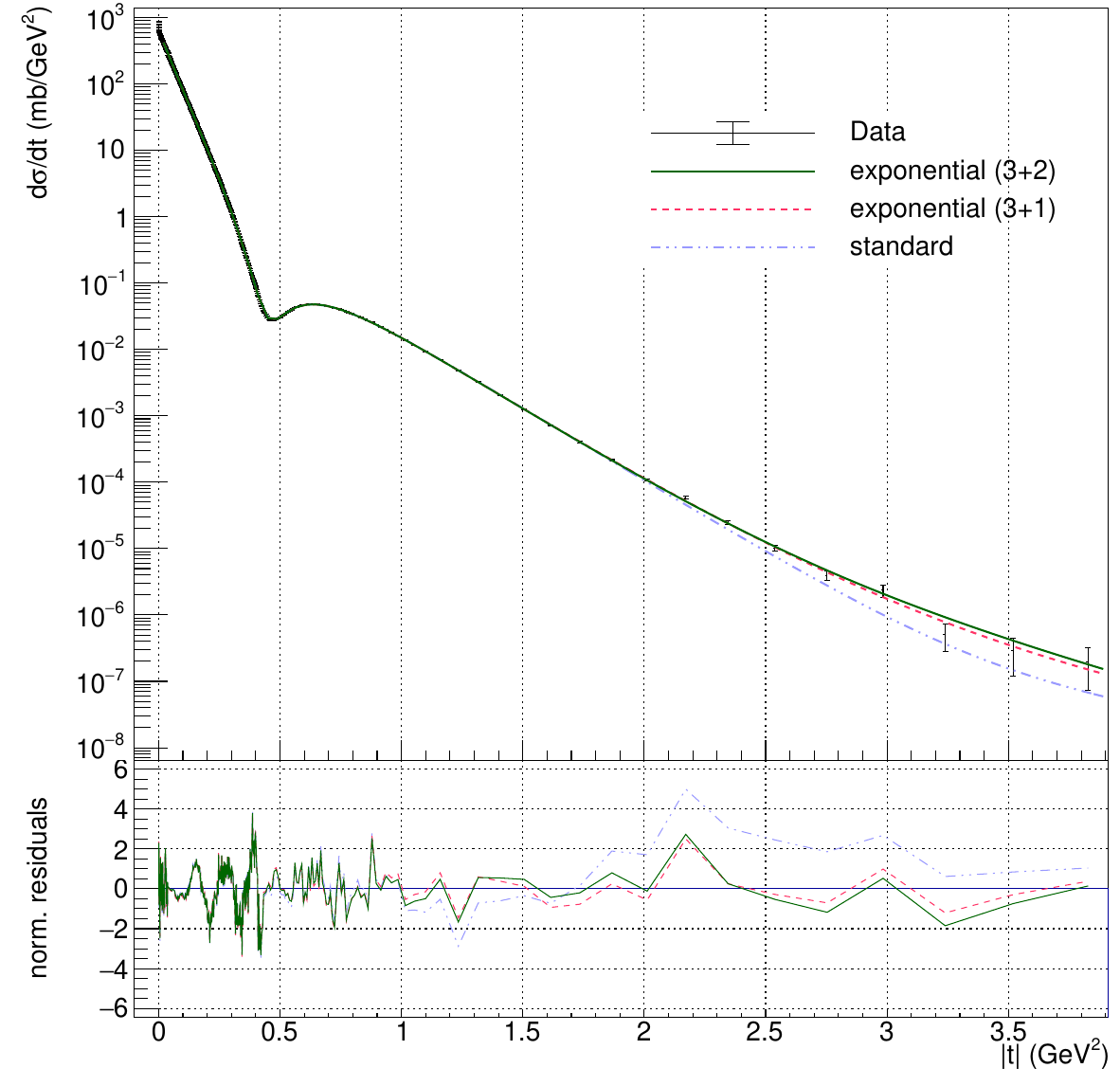}
	\caption{\label{fig:fit-quality} TOTEM data \cite{Antchev:2017yns,Ravera:2018,*Nemes:2018} description (color online) by the parameterizations \eqref{eq:standard-parameterization} and two variants of \eqref{eq:generic-parameterization} (upper panel) and corresponding normalized residuals \eqref{eq:residuals} (lower panel).}
\end{figure}
\paragraph{Fit quality.} Fit quality is assessed by distribution of normalized residuals defined as
\begin{equation}\label{eq:residuals}
r(t_i) = \frac{\left. d\sigma/ dt \right|_{t_i} - \kmb / \left(16\pi s^2\right) \left| A^{\mathrm{param}}(s,t_i) \right|^2}{\left.\delta_{d\sigma/dt}\right|_{t_i}},
\end{equation}
where $A^{\mathrm{param}}$ is the particular parameterization and $\left.\delta_{d\sigma/dt}\right|_{t_i}$ is the experimental uncertainty of differential cross-section data excluding the normalization uncertainty.
Fits are presented visually together with the residuals distribution in Fig.~\ref{fig:fit-quality}.
In summary, exponential (3+1) provides the best description for differential cross-section data at 13~TeV, closely followed by exponential (3+2).
While standard parameterization is less successful, it is still used to compare with previous results and to provide a different $\Re A(t)$ behavior to assess the model dependence of the $H(b)$ extraction method.
Best-fit parameter values and corresponding $\chi^2$ values are given in the Table \ref{tab1}.
\paragraph{\label{par:ImH}Imaginary part of the impact-parameter amplitude\footnote{T. Fearnley \cite{Fearnley:1985uy} used the profile function $\Gamma(s,b)=-2iH(s,b)$}.} Imaginary part of the elastic amplitude is directly calculated, following \eqref{eq:b-transform-to}, as
\begin{equation}\label{eq:bins1}
 \Im H^{\mathrm{(d)}}(b)=\frac{1}{8\pi s} \int_{0}^{q_{\mathrm{max}}} \dqq
\bess \Im A_{N}(q),
\end{equation}
where $A_{N}(s,t)$ is the nuclear hadronic amplitude (without Coulomb contribution) taken from a parameterization and $q_{\mathrm{max}} \equiv \sqrt{\left|t_{\mathrm{max}}\right|}$ with $\left|t_{\mathrm{max}}\right| \approx 5$~GeV$^2$.
It can be explicitly shown that the result is not sensitive to increase in $|t_{\mathrm{max}}|$ if it is large enough and does not depend on the particular $A(t)$ function, provided it reasonably extrapolates to large $|t|.$
We emphasize here, that we have discarded the calculation method, originally used by Amaldi et al. \cite{Amaldi:1979kd} and Fearnley \cite{Fearnley:1985uy} and that we had previously applied to 7~TeV data \cite{Alkin:2014rfa}. 
The naive error propagation based on that method was found to be unreliable and improvement of it would be unnecessary complex. 
Moreover, our own application of the method was flawed, which lead to a fallacious conclusion of black disk limit excess for 7~TeV data \cite{Alkin:2014rfa}. 

\paragraph{Real part of the impact-parameter amplitude.} Real part of the elastic amplitude in impact-parameter representation is calculated directly using the transformation \eqref{eq:b-transform-to}
\begin{equation}
\Re H(b) = \frac{1}{8\pi s}
\int_{0}^{q_{\mathrm{max}}}\dqq \bess \Re A(q).
\end{equation}
Finally, we can determine inelastic overlap function\footnote{The inelastic overlap function alone is not enough to draw conclusions about asymptotic regime as shown in previous analyses \cite{Amaldi:1979kd, Fearnley:1985uy,Dremin:2013qua}} $\Ginel(s,b)$ using unitarity condition in impact-parameter representation at the high-energy limit
\begin{equation}\label{eq:unitarity}
\Im H(s,b)=|H(s,b)|^{2}+\Ginel(s,b).
\end{equation}
\paragraph{Uncertainty calculation.} As the quantities under consideration depend on the data non-trivially, uncertainties from the experimental points were propagated numerically by varying those within their respective limits (assuming the quoted uncertainty to be $1\sigma$  interval) producing a corresponding set of results for $\Im H(b)$ and $\Re H(b)$. 
A sample of three hundred varied data sets was produced.
For each data set, values $\Im H(b)$ and $\Re H(b)$ were calculated using the procedure described above, starting with the fit of differential cross-section data.
At each value of impact parameter $b$, the final values of amplitude imaginary and real parts, $\Im H(b)$ and $\Re H(b)$, were calculated as the sample average with the corresponding uncertainty given by standard deviation. 
This propagation procedure was extensively tested for stability and robustness.
Central values of imaginary and real parts of the amplitude are independent of sample size, given it is large enough, and are completely independent of starting parameter values for differential cross-section data fits. 
An increase in input uncertainty produces a proportional increase in the uncertainties of final quantities.
We have specifically confirmed, that distributions of sampled $\Im$ and $\Re H$ values at each $b$ are roughly normal and do not contain significant outliers.
Exponential (3+2) parametrization has slightly lesser stability and the distributions of $\Im$ and $\Re H$ samples are wider than those for standard and exponential (3+1) parametrization, and and deviate from normal, which is reflected in larger uncertainty of the final values obtained with this parameterization (see Fig. \ref{fig:im-h-scaled-0}).
Uncertainty of inelastic overlap $\Ginel(b)$ was calculated from uncertainties of $\Im$ and $\Re H(b)$ using the standard propagation procedure.

Additionally, the normalization uncertainty of the differential cross-section data is taken into account.
Starting with the expression \eqref{eq:normalization}, assuming the relative normalization uncertainty of $d\sigma/dt$, $\varepsilon_{d\sigma/dt} \approx 0.03$~\cite{Ravera:2018,*Nemes:2018}, we can write
\begin{multline}
(1 \pm \varepsilon_{d\sigma/dt})\frac{d\sigma}{dt} = \frac{\kmb}{16\pi s^2}  \left|\left(1 \pm \varepsilon_{A}\right) A(s,t)\right|^2 \\
 \approx \frac{\kmb}{16\pi s^2}  \left(1 \pm 2\varepsilon_{A}\right) \left|A(s,t)\right|^2,
\end{multline}
where $\varepsilon_{A}$ is the corresponding relative uncertainty on scattering amplitude. 
Thus, we can derive that $\varepsilon_{A} = 1/2\ \varepsilon_{d\sigma/dt}$.
There is still an ambiguity on assigning this relative uncertainty individually to real and imaginary parts of the amplitude.
To avoid unnecessary complications, we assume that $\varepsilon_{A}$ affects both real and imaginary parts of the amplitude similarly, thus $\varepsilon_{\Im A} = \varepsilon_{\Re A} = \varepsilon_{A} = 1/2\ \varepsilon_{d\sigma/dt}.$
Additionally, due to constraints \eqref{eq:constraints}, amplitude uncertainty already partially includes the normalization uncertainty coming from a direct dependence on $\sigmat$ and $\rho$, however accounting for that would be an overcomplication.
The relative normalization uncertainty propagates with no modification into $\Im$ and $\Re H$, $\varepsilon_{\Im H} = \varepsilon_{\Re H} = 1/2\ \varepsilon_{d\sigma/dt}.$ 

Finally, it is propagated to $\Ginel$ by standard methods, accounting for the fact that it is fully correlated between $\Im$ and $\Re H$
\begin{equation}
\varepsilon_{\Ginel} =  \left| 1 - \frac{\Im H(b)}{2\Ginel(b)} \right| \varepsilon_{d\sigma/dt}.
\end{equation}
Note that normalization uncertainty on $\Ginel$ is anticorrelated o that of $\Im$ and $\Re H$ for $b \lesssim 0.5$~fm.

\begin{table*}
	\centering
	\caption{Best-fit parameters for parameterizations \eqref{eq:standard-parameterization} and two variants of \eqref{eq:generic-parameterization} using data at 13~TeV \cite{Antchev:2017yns,Ravera:2018,*Nemes:2018}.}
	\label{tab1}
	\begin{ruledtabular}
		\def\arraystretch{1.2}%
	\begin{tabular}{@{}c c d@{}d<{\mkern-20mu}@{}d d@{}d<{\mkern-20mu}@{}d d@{}d<{\mkern-20mu}@{}d@{}}
		                 \multicolumn{2}{c}{}                  &                                                 \multicolumn{9}{c}{\textbf{Parameterization}}                                                  \\
	\cline{3-11}
		        \textbf{Parameter}         & \textbf{Unit}     &    \multicolumn{3}{c}{\textbf{standard}}     & \multicolumn{3}{c}{\textbf{exponential (3+1)}} & \multicolumn{3}{c}{\textbf{exponential (3+2)}} \\
	\hline
		            $\bm{A_1}$             & none / GeV$^{-2}$ & 0.335  & \pm & 0.005                         & -17.0 & \pm & 0.3                          & 197.3 & \pm & 2.0                              \\
		            $\bm{A_2}$             & GeV$^{-2}$        & 17.91  & \pm & 0.09                          & 117.0  & \pm & 2.5                           & -19.1 & \pm & 0.2                              \\
		            $\bm{A_5}$             & GeV$^{-2}$        & \multicolumn{3}{c}{\text{\hspace{3.2ex}N/A}} &         \multicolumn{3}{c}{\text{N/A}}         & -12.7  & \pm & 0.4                             \\
		            $\bm{A_p}$             & GeV$^{-2}$        & \multicolumn{3}{c}{\text{\hspace{3.2ex}N/A}} & -13.4  & \pm & 3.1                           &         \multicolumn{3}{c}{\text{N/A}}         \\
		            $\bm{b_1}$             & GeV$^{-2}$        & 0.0956 & \pm & 0.0006                        & 2.63   & \pm & 0.02                          & 7.80  & \pm & 0.03                             \\
		            $\bm{b_2}$             & GeV$^{-2}$        & 0.0517 & \pm & 0.0001                        & 14.04   & \pm & 0.09                           & 2.61  & \pm & 0.01                              \\
		            $\bm{b_3}$             & GeV$^{-2}$        & 5.041  & \pm & 0.008                         & 7.67   & \pm & 0.03                          & 14.2  & \pm & 0.1                             \\
		            $\bm{b_4}$             & GeV$^{-2}$        & \multicolumn{3}{c}{\text{\hspace{3.2ex}N/A}} & 7.112   & \pm & 0.003                          & 17.9  & \pm & 5.0                              \\
		            $\bm{b_5}$             & GeV$^{-2}$        & \multicolumn{3}{c}{\text{\hspace{3.2ex}N/A}} &         \multicolumn{3}{c}{\text{N/A}}         & 5.00  & \pm & 0.05                             \\
		           $\bm{\tau}$             & GeV$^{2}$         & 0.942  & \pm & 0.009                         & 0.56   & \pm & 0.04                          &         \multicolumn{3}{c}{\text{N/A}}         \\
	\hline
		          $\bm{\sigmat}$           & mb                &             \multicolumn{3}{c}{}             & 112.05  & \pm & 0.05                           &              \multicolumn{3}{c}{}              \\
		           $\bm{\rho}$             & none              &             \multicolumn{3}{c}{}             & 0.099   & \pm & 0.001                          &              \multicolumn{3}{c}{}              \\
	\hline
		$\bm{\chi^2/\text{\bfseries NDF}}$ &                   & 1.277  &        \multicolumn{2}{c}{}         & 1.097   &         \multicolumn{2}{c}{}         & 1.109 &          \multicolumn{2}{c}{}
	\end{tabular}
	\end{ruledtabular}
\end{table*}
\section{\label{sec:results}Results}
\begin{figure}[t]
	\subfloat[\label{fig:main-result-overall} $H(s,b)$, $\Ginel(s,b)$]
	{
		\includegraphics[width=\linewidth]{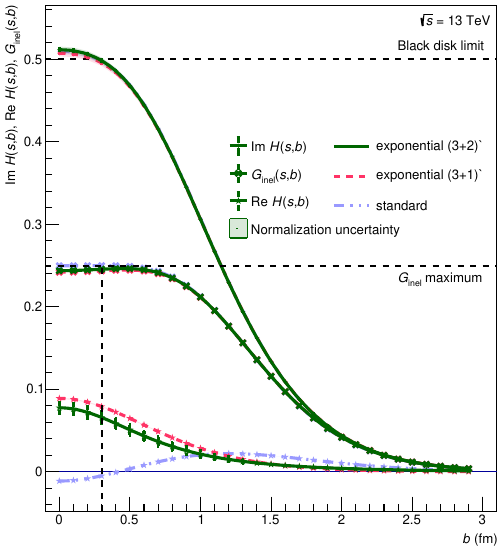}%
	} \\%
	\subfloat[\label{fig:g-inel-scaled-max} $\Ginel(s,b)$]
	{
		\includegraphics[width=\linewidth]{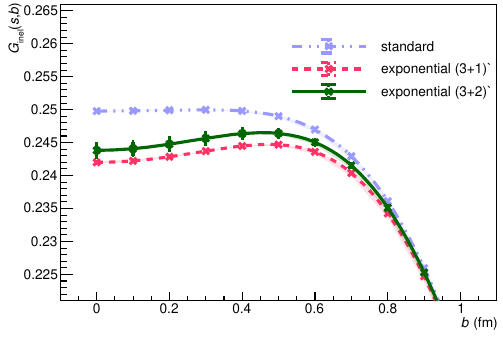}%
	}\\
	\caption{\label{fig:results}Graphic representation of analysis results (color online): \protect\subref{fig:main-result-overall} real and imaginary parts of elastic amplitude $H(s,b)$ and inelastic overlap $\Ginel(s,b)$ as functions of impact parameter $b$; \protect\subref{fig:g-inel-scaled-max} scaled up region of $\Ginel(s,b)$ maximum.}
\end{figure}
We have considered three parameterizations, \eqref{eq:standard-parameterization} and two variants of \eqref{eq:generic-parameterization}, for pp elastic scattering amplitude at  $\sqrt{s}=13$~TeV, in order to assess the relative importance of elastic amplitude real part $\Re A(s,t)$ for $H(s,b)$ reconstruction and the overall sensitivity of the process to the particular functional form of the real part.

Reconstructed functions $H(s,b)$ and $\Ginel(s,b)$ are presented in Fig. \ref{fig:main-result-overall}.
Figure \ref{fig:g-inel-scaled-max} shows a scaled-up version of $\Ginel(s,b)$ plot near the maximum.

The behavior of imaginary part of the amplitude as a function of $t$ is practically identical in three parameterizations used, since it is mostly determined by the differential cross-section data.
Only considerable differences can be observed near the dip region.
Real part, however, is less constrained, which allows us to assess its effect on the final quantities.
Figure \ref{fig:main-result-overall} shows that in impact-parameter representation real part of the amplitude is different between the parameterizations, but the effect on the imaginary part behavior is negligible.

\begin{figure}[t]
	\subfloat[\label{fig:im-h-scaled-0}][$\Im H(s,b)$ at 13~TeV]
	{
		\includegraphics[width=\linewidth]{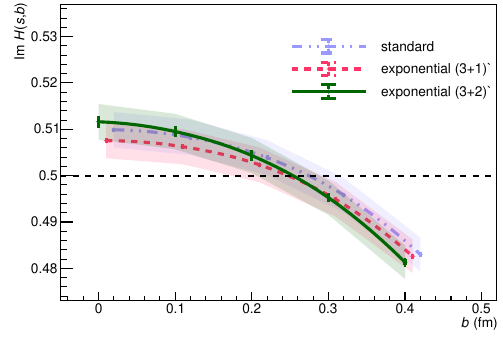}
	}\\
	\subfloat[\label{fig:im-h-extrapolation}][$\Im H(s,0)$ as a function of $s$]
	{
		\includegraphics[width=\linewidth]{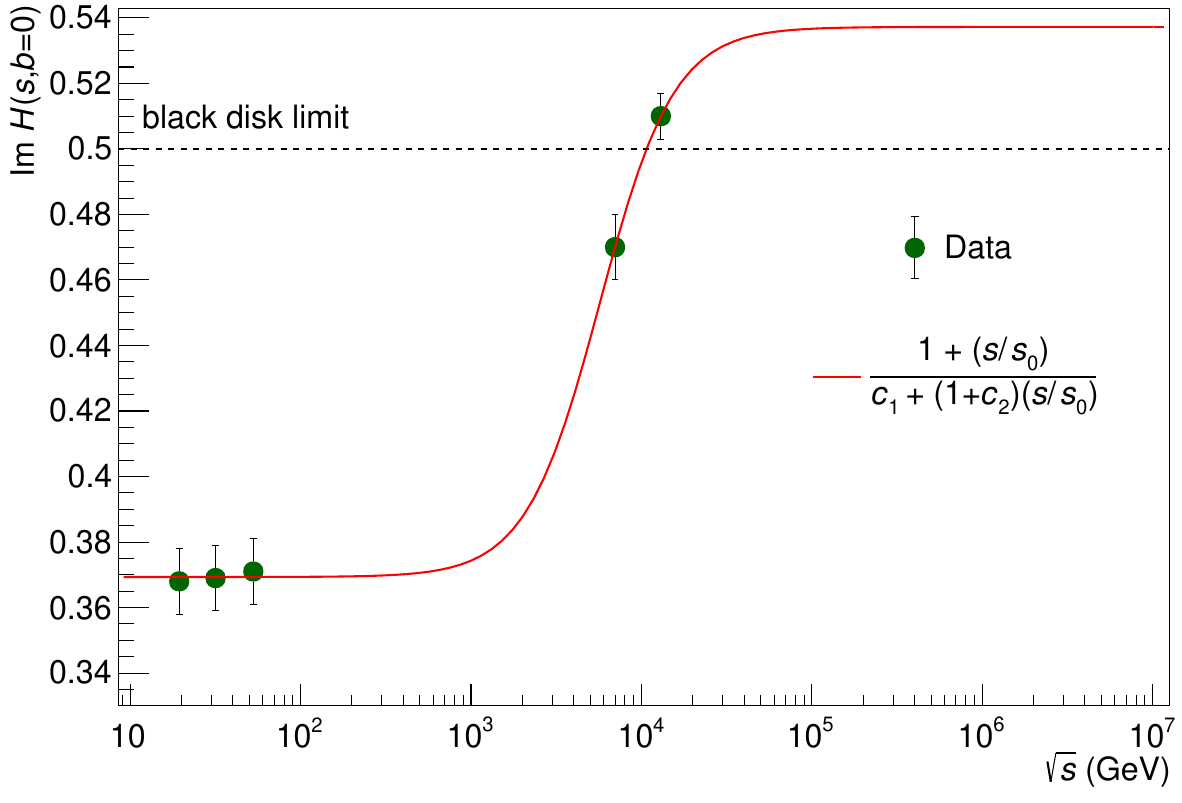}
	}
	\caption{\label{fig:im-h-full}\protect\subref{fig:im-h-scaled-0} Scaled-up version $\Im H(s,b)$ near $b=0$ (for each of the curves markers are placed at slightly different values of $b$ for clarity); \protect\subref{fig:im-h-extrapolation} extrapolation of $\Im H(s,0)$ to higher energies.}
\end{figure}

The main conclusion from comparison of the different amplitude parameterizations is the following.
From Figures \ref{fig:main-result-overall} and \ref{fig:g-inel-scaled-max} we can see that black disk limit is exceeded in pp collisions at 13~TeV. 
Imaginary part of the amplitude at $b=0$ exceeds $1/2$ at 13~TeV, but not at 7~TeV (it was claimed to be already observed at 7~TeV \cite{Alkin:2014rfa}, but the analysis was flawed). 
Values, reached with the three parameterizations considered, are $\Im H(s,b) = 0.512\pm 0.001(\text{sys+stat})\pm 0.004(\text{norm})$ (exponential (3+2)), $0.5076\pm 0.0002(\text{sys+stat})\pm 0.0038(\text{norm})$ (exponential (3+1)) and $0.5099\pm 0.0001(\text{sys+stat})\pm 0.0038(\text{norm})$ (standard).
Figure \ref{fig:im-h-scaled-0} shows the $\Im H(s,b)$ behavior in $b \approx 0$ region in a larger scale.

We would like to note that the first results on $H(s,b)$ at $\sqrt{s}=13$~TeV were presented at the 4$^{\text{th}}$ Elba Workshop on Forward Physics @ LHC Energy in the talks of E.~Martynov and A.D.~Martin \cite{Martynov:2018,*Martin:2018}, albeit without error analysis.

Recently some similar estimates  were published \cite{Khoze:2018kna}, also without an error analysis.
It is claimed that effect is too small and the value ''is consistent with the statement that the amplitude \emph{does not exceed black disk limit}``.
However, taking into account the error analysis we have performed here, we cannot support the statement that impact-parameter amplitude conforms to the black disk limit.
Additional information can be gained from investigating the growth trend of $\Im H(0)$ value with energy.
The data at lower energies (the same data set that was used previously \cite{Alkin:2014rfa}) was re-analyzed to extract values of $\Im H(s,0)$ and the average is taken between three parameterizations. 
The results are plotted as a function of $s$ in Fig. \ref{fig:im-h-extrapolation}.
Data points are fitted with a simple function to extrapolate the behavior at higher energies.
The function was chosen to have an asymptotic limit lower than 1 at $s\to\infty$ and to conform to the generic features of data, such as apparent flatness at few tens of GeV and a rapid growth at TeV energies.
The minimal growth is given by rational function is used in a form
\begin{equation}
	F(s) = \frac{1 + \left(s/s_0\right)}{c_1 + \left(1+c_2\right)\left(s/s_0\right)},
\end{equation}
where $c_i$ and $s_0$ are free parameters, $c_2 \geq 0$.
The growth trend between 7 and 13~TeV implies that possible asymptotic values of the impact-parameter amplitude exceed $1/2$, even if we assume that current analysis overestimates the value of $\Im H(0)$ at 13~TeV.
Function of $\ln^{p} s$ was also considered, however it requires $p \approx 6$ to fit the growth at TeV energies.
The most pessimistic extrapolation with asymptotic value fixed at $1/2$ was found to be not compatible with the data.
More points at intermediate energies (from few hundred GeV to 2--3 TeV) would be most useful to better fix the trend. 


\section{Conclusion}
The black disk limit excess leads to unitarity saturation characterized by reflective scattering mode dominance \cite{Troshin:2007fq}.
Its main feature is a negativity of the elastic scattering matrix element $S(s,b)$ (where $b$ is an impact parameter of the colliding hadrons; note that angular momentum $l \sim b\sqrt{s}/2$) leading to the asymptotic dominance of the reflective elastic scattering  and peripheral form of the inelastic overlap as a function of the impact parameter.
The corresponding elastic scattering decoupling from the multiparticle production occurs initially at small values of the impact parameter $b$ expanding to larger values with increase of energy. 
Such behavior corresponds to increasing self-dampening of inelastic contributions to unitarity equation \cite{Baker:1962zza}.

The $b$-dependence of the scattering amplitude as well as the inelastic overlap function should be considered as a collision geometry.
It should be emphasized that the collision geometry describes the hadron interaction region but not the matter distribution inside of the individual colliding hadrons.

For qualitative discussion it is convenient to assume smallness of the real part of the elastic scattering amplitude in the impact parameter representation $H(s,b)$ and substitute $H\to iH$. 
This assumption is related to unitarity saturation, meaning that, at $s\to\infty$ with $b$, $\Im H(s,b)\to 1$ and $\Re H(s,b)\to 0$. 
However, alternatives exist where $\Im H(s,b)\to H_{0}>1/2$ but $\Re H(s,b)\nrightarrow 0$ at $s\to\infty$.
It is claimed \cite{Petrov:2018wlv} that accounting the real part of the elastic scattering amplitude $H$ should lead to the central dependence of $\Ginel$ on $b$.
However, present analysis demonstrates that peripheral behavior of $\Ginel$ is still observed even when the real part of the impact-parameter amplitude is non-negligible.
It is also important to note that peripheral mode is achieved without $\Ginel$ reaching unitarity limit of $1/4$.
This is especially visible with the parameterizations introduced in present analysis.

The present analysis supersedes results of the previous one performed for lower energy of 7~TeV.
We demonstrate that the elastic scattering amplitude slightly exceeds the black disk limit at $\sqrt{s}=13$, and that inelastic overlap function is peripheral.
Results are also consistent with smooth energy dependence of the elastic scattering amplitude at LHC energies and one can conclude that observed growth of $\Im H(s,0)$ with energy indirectly supports the conclusion that black disk limit is violated.

We would like to note that the important consequence of black disk limit excess, if it is fully confirmed, is that the pp and $\overline{\text{p}}$p scattering models, based on eikonal approach, must be substantially modified to be fundamentally compatible with experimental data (one of the recent studies demonstrate that eikonal fails to describe differential cross-section data \cite{Durand:2018irx}). 

\begin{acknowledgments}
We are grateful to N.E.~Tyurin, V.A.~Petrov, J.~Kaspar and K. \"{O}sterberg for the fruitful and interesting discussions on various aspects.  
The present work was partially supported by the Program of Fundamental Research of the Department of Physics and Astronomy of the National Academy of Sciences of Ukraine  (project No. $0117$U$000240$).
\end{acknowledgments}
%

\end{document}